\documentclass[aps,prb,twocolumn,superscriptaddress,10pt]{revtex4-1}
\usepackage{graphicx,amsmath,amssymb,amsfonts}

\begin{document}


\title{All-electrical coherent control of the exciton states in a single quantum dot}


\author{A. Boyer de la Giroday}
\email[Electronic address: ]{\texttt{ab783@cam.ac.uk}}
\affiliation{Toshiba Research Europe Limited, Cambridge Research Laboratory, 208 Science Park, Milton Road, Cambridge, CB4 0GZ, U. K.}
\affiliation{Cavendish Laboratory, Cambridge University, JJ Thomson Avenue, Cambridge, CB3 0HE, U. K.}
\author{A.J. Bennett}
\affiliation{Toshiba Research Europe Limited, Cambridge Research Laboratory, 208 Science Park, Milton Road, Cambridge, CB4 0GZ, U. K.}
\author{M.A. Pooley}
\affiliation{Toshiba Research Europe Limited, Cambridge Research Laboratory, 208 Science Park, Milton Road, Cambridge, CB4 0GZ, U. K.}
\affiliation{Cavendish Laboratory, Cambridge University, JJ Thomson Avenue, Cambridge, CB3 0HE, U. K.}
\author{R.M. Stevenson}
\affiliation{Toshiba Research Europe Limited, Cambridge Research Laboratory, 208 Science Park, Milton Road, Cambridge, CB4 0GZ, U. K.}
\author{N. Sk\"{o}ld}
\affiliation{Toshiba Research Europe Limited, Cambridge Research Laboratory, 208 Science Park, Milton Road, Cambridge, CB4 0GZ, U. K.}
\author{R.B. Patel}
\affiliation{Toshiba Research Europe Limited, Cambridge Research Laboratory, 208 Science Park, Milton Road, Cambridge, CB4 0GZ, U. K.}
\affiliation{Cavendish Laboratory, Cambridge University, JJ Thomson Avenue, Cambridge, CB3 0HE, U. K.}
\author{I. Farrer}
\affiliation{Cavendish Laboratory, Cambridge University, JJ Thomson Avenue, Cambridge, CB3 0HE, U. K.}
\author{D.A. Ritchie}
\affiliation{Cavendish Laboratory, Cambridge University, JJ Thomson Avenue, Cambridge, CB3 0HE, U. K.}
\author{A.J. Shields}
\affiliation{Toshiba Research Europe Limited, Cambridge Research Laboratory, 208 Science Park, Milton Road, Cambridge, CB4 0GZ, U. K.}

\date{\today}

\begin{abstract}
We demonstrate high-fidelity reversible transfer of quantum information from the polarisation of photons into the spin-state of an electron-hole pair in a semiconductor quantum dot. Moreover, spins are electrically manipulated on a sub-nanosecond timescale, allowing us to coherently control their evolution. By varying the area of the electrical pulse, we demonstrate phase-shift and spin-flip gate operations with near-unity fidelities. Our system constitutes a controllable quantum interface between flying and stationary qubits, an enabling technology for quantum logic in the solid-state.
\end{abstract}


\maketitle



Emerging quantum technologies such as quantum computing and quantum cryptography promise to revolutionise the way information is processed by controlling quantum bits or ``qubits'' in two-level quantum systems \cite{nielsen,bouwmeester}. Photonic ``flying'' qubits are used for quantum communication \cite{gisin,zeilinger2,zeilinger1} as they easily travel in space or through standard optical fibres while solid-state ``stationary'' qubits are usually more convenient to perform quantum computations \cite{ladd,koppens,press,greilich}. These two embodiments have thus far been developed independently, but coherent control of both types of qubits and coherent transfer of quantum information between them are necessary to extend the potential of quantum technologies \cite{kosaka}.

Optical absorption of photons in a semiconductor to create an electron-hole pair (or exciton) provides a natural mechanism to initialise a stationary qubit. Selection rules determine the mapping of the photon polarisation onto the spin-state of the exciton. Once initialised manipulation of solid-state spins has been achieved with pulsed magnetic fields at milli-kelvin temperatures \cite{koppens}, coherent optical beams \cite{press,greilich} or couplings to localised optical modes \cite{imamoglu}. However, these techniques require bulky setups or synchronised lasers, making them somewhat impractical and incompatible for large-scale applications. New methods to control these qubits that take full advantage of well-established semiconductor technology will be advantageous when it comes to controlling large numbers of qubits.

Here, we advocate using the spin-state of an exciton trapped in a self-assembled semiconductor quantum dot (QD) as an electrically-controllable qubit in which quantum information from flying qubits can be efficiently transferred and easily restored. An exciton has two optically-active spin eigenstates which couple to linearly polarised photons and are energetically separated by a fine-structure splitting ($s$) resulting from the exchange interaction \cite{bayer,gammon}. We control the splitting and the orientation of the eigenstates via an applied vertical electric field \cite{bennett}, which allows for sub-nanosecond electrical coherent spin manipulation. We demonstrate high-fidelity phase-shift and spin-flip operations using dynamical modulation of the electric field.

The device used was similar in design to those of previous experiments \cite{patel} and consists in a $p-i-n$ heterostructure grown by molecular beam epitaxy. A single layer of InAs QDs with dot density $<$1$\mu$m$^{-2}$ was grown at the center of the intrinsic region made of a 10nm GaAs quantum well clad with a short period superlattice equivalent to Al$_{0.75}$Ga$_{0.25}$As on each side, which prevents tunneling of the carriers out of the dot region when the structure is biased. Doping extends into the superlattice and allows application of an electric field along the growth direction. The electric field $F$ is calculated using $F=\left(\frac{V-V_{bi}}{d}\right)$, where $V$ is the bias applied to the structure, $V_{bi}$=2.2V is the built-in potential and $d$=140nm is the thickness of the intrinsic region. This $p-i-n$ device is encased in a weak planar microcavity consisting of 14 (4) periods below (above) the dot layer. We used standard photolithography and wet etching techniques to fabricate a diode with an area of 35$\mu$m$\times$60$\mu$m. Excitation and photon collection occurs through an opaque metallic film on the sample surface patterned with micron-diameter apertures. The sample was cooled to $\approx$5K and excited quasi-resonantly using a mode-locked Ti:Sapphire laser.

We start by characterising the physical properties of our QD used as a solid-state photonic interface. Fig. 1(a) shows a schematic energy diagram of the QD. The exciton spin-state is initialised through quasi-resonant optical excitation and phonon relaxation. The phonon energy was measured to be $\approx$34.6meV corresponding to the 1-LO phonon in bulk GaAs. The eigenstates $\left|X_{H}\right\rangle$ and $\left|X_{V}\right\rangle$, coupled to linearly polarised photons, are separated by a tunable energy $s$. A vertical electric field reduces $s$, eventually causing an inversion in its sign. In Fig. 1(b), $\left|s\right|$ is shown to vary linearly with gradient $\pm$0.26$\mu$eVkV$^{-1}$cm. An anticrossing is measured at a field $F_{0}$=-155.4kVcm$^{-1}$ where the splitting is reduced to its minimum value of $s_{0}$=0.4$\mu$eV. This anticrossing originates from a field-dependent coupling between the eigenstates.

This coupling also leads to mixing of the eigenstates polarisation character and thus rotation of their orientation (Fig. 1(c)). We use the notation $\left|X_{H}\right\rangle$ and $\left|X_{V}\right\rangle$ to refer to the spin eigenstates away from the anticrossing, where $H$ and $V$ are the horizontal and vertical orientations in the lab-frame defined by the polarisation of photons to which the eigenstates couple. As the field is swept across $F_{0}$, the polarisation of those photons rotates by 90 degrees. At the anticrossing, the eigenstates $\left|X_{D}\right\rangle$ and $\left|X_{A}\right\rangle$ couple to photons with diagonal ($D$) and anti-diagonal ($A$) polarisations in the lab-frame.

Importantly, quasi-resonant optical excitation allows for initialisation of any superposition of spin-states by mapping the polarisation of an ``input'' excitation photon into the spin-state of the exciton \cite{flissikowski,kowalik}. Considering an input photon of polarisation $\left|\Psi_{in}\right\rangle=\cos\theta\left|H\right\rangle+e^{i\varphi}\sin\theta \left|V\right\rangle$, the idealised coherent time evolution of the exciton spin-state (away from the anticrossing) is given by

\begin{equation}
\left|\Psi_{X}(t)\right\rangle=\cos\theta\left|X_{H}\right\rangle+e^{i\frac{st}{\hbar}}e^{i\varphi}\sin\theta\left|X_{V}\right\rangle.
\end{equation}

The splitting $s$ therefore introduces a phase difference between the eigenstates accumulated over time at a rate of $st/\hbar$ and leading to coherent oscillations of the spin-state \cite{hudson}. Those oscillations can be observed by initialising an exciton and measuring its radiative emission in superpositions of eigenstates, leading to an ideal signal of intensity $I(t)\propto\cos\left(st/\hbar\right)e^{-t/\tau_{r}}$ (Fig. 1(d)), where $\tau_{r}$ is the radiative lifetime measured to be 1.28ns$\pm$0.08ns for our QD. Fitting those oscillations allows for temporal measurements of $\left|s\right|$ that agree well with spectral measurements (Fig. 1(b)).

\begin{figure}
\includegraphics[width=1.\columnwidth,keepaspectratio]{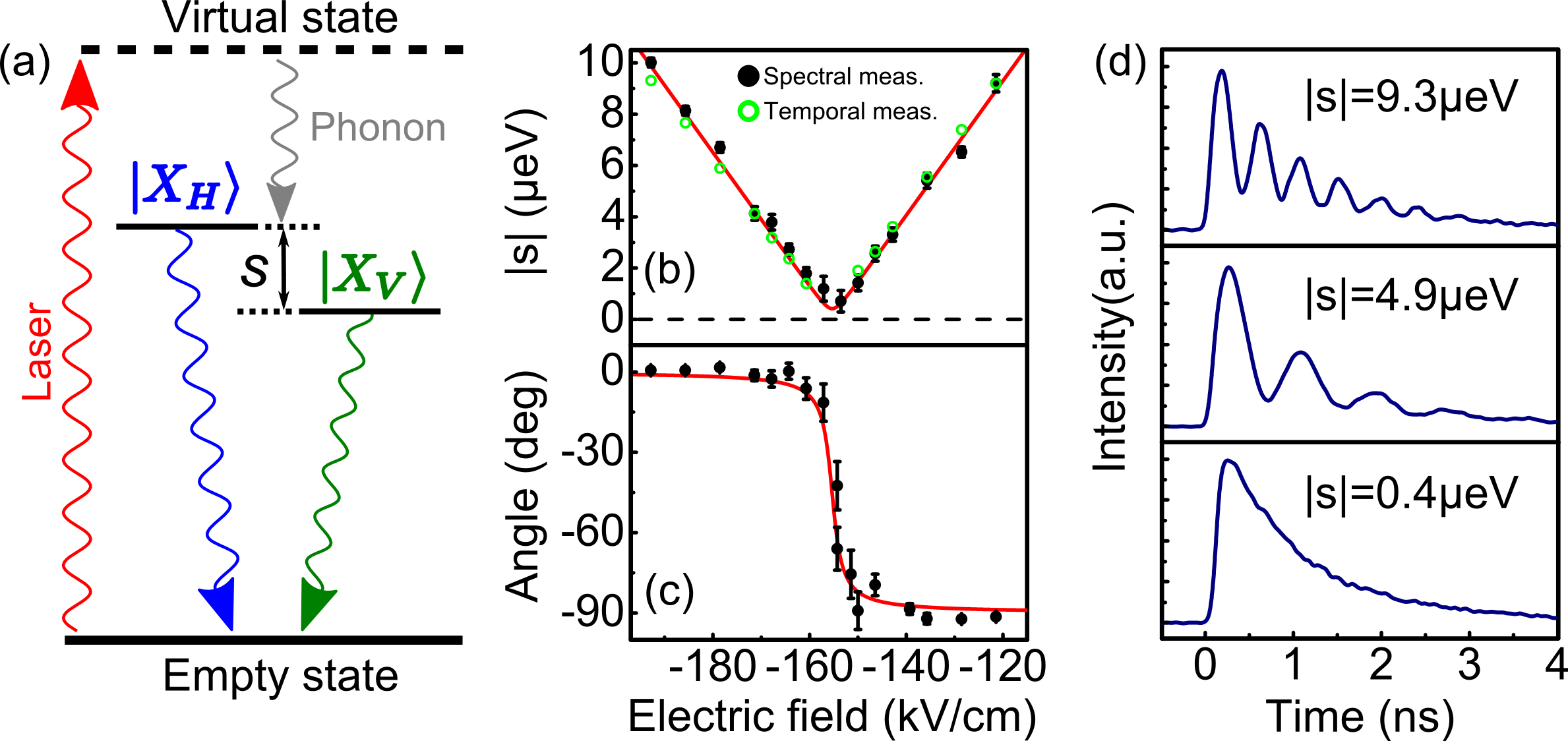}%
\caption{(colour online). (a) Schematic energy diagram of the QD used in our experiment where the polarisation of the laser is transferred into the exciton's spin-state through quasi-resonant excitation. (b) and (c) Evolution of $\left|s\right|$ and orientation of the eigenstates relative to the lab-frame as a function of vertical electric field. Spectral and temporal measurements of $\left|s\right|$ are in good agreement. (d) High-visibility coherent oscillations of the spin-state for three different values of $\left|s\right|$. The exciton was initialised in a maximum superposition of eigenstates and emission was time-resolved along the same orientation.\label{fig1}}
\end{figure}

We now characterise the use of an exciton as an interface between optical and solid-state qubits (Fig. 2). Measurements were performed at an electric field of -175kVcm$^{-1}$ where $\left|s\right|\approx$5$\mu$eV and where the eigenstates are oriented along $H$ and $V$. Fig. 2(a) illustrates the operating principle where an input qubit $\left|\Psi_{in}\right\rangle$ is first encoded into the polarisation of a photon defined by the parameters $\theta$ and $\varphi$ that are used as coordinates to represent the optical qubit on the Poincar\'{e} sphere. Quasi-resonant excitation of the QD then maps the polarisation of $\left|\Psi_{in}\right\rangle$ into the spin-state of the exciton qubit $\left|\Psi_{X}(t)\right\rangle$, therefore defining a point with the same coordinates ($\theta$,$\varphi$) on the Bloch sphere of the solid-state qubit. The phase accumulation between the eigenstates rotates the qubit around the equator of the Bloch sphere. Eventually, the electron-hole pair in the QD recombines and emits a photon of polarisation $\left|\Psi_{out}\right\rangle$ depending on the time spent in the solid-state $\left|\Psi_{X}(t)\right\rangle$.

The time evolution of $\left|\Psi_{X}(t)\right\rangle$ is characterised in Fig. 2(b), where the exciton was initialised in a diagonal superposition of eigenstates (corresponding to $\left|X_{D}\right\rangle$ on the Bloch sphere) and the emission was measured in linear $\left\{\left|H\right\rangle,\left|V\right\rangle\right\}$, diagonal $\left\{\left|D\right\rangle,\left|A\right\rangle\right\}$ and circular $\left\{\left|R\right\rangle,\left|L\right\rangle\right\}$ bases. According to Eq. (1) and Fig. 2(a), no oscillations are observed when measuring $\left|\Psi_{out}\right\rangle$ along the eigenstates while oscillations of angular frequency $s/\hbar$ are observed when measuring in diagonal and circular basis states, with a dephasing corresponding to the respective positions of those states on the equator.

We measure the fidelity of the interface $f_{in}$=$\left|\left\langle\Psi_{out}|\Psi_{in}\right\rangle\right|^{2}$ for six different input qubits prepared in the linear, diagonal, and circular states (Fig. 2(c)). Exciting an eigenstate leads to $f_{in}$=0.95$\pm$0.03 decaying with a timescale of 78ns$\pm$17ns, which is limited by spin scattering. Exciting a maximum superposition leads to an initial fidelity $f_{in}$=0.81$\pm$0.03 which oscillates due to the finite value of $s$. The envelope of the time-resolved fidelity decays with a timescale of 3.0ns$\pm$0.4ns limited by cross-dephasing, which randomizes the phase relationship between the two superimposed eigenstates \cite{stevenson2,hudson}. Cross-dephasing is significantly longer than the radiative lifetime and the time needed to perform electrical manipulation of the qubit. The amplitude of the oscillation in the fidelity is thought to be reduced by the finite response time of the APD as well as the uncertainty in the time at which the superposition is created originating from the finite linewidth of the transition.

\begin{figure}
\includegraphics[width=1.\columnwidth,keepaspectratio]{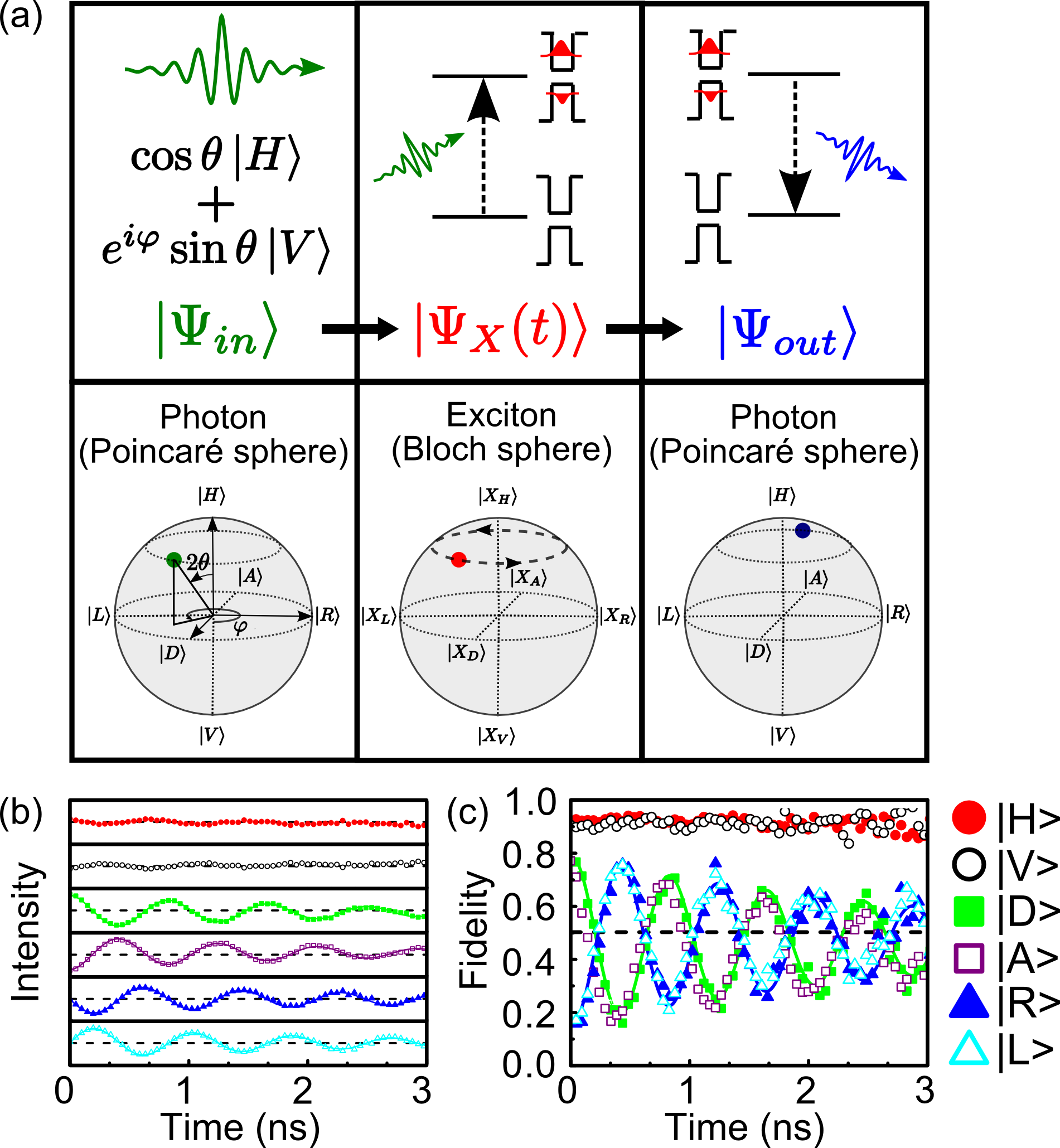}
\caption{(colour online). (a) Operating principle of our solid-state photonic interface. An optical qubit $\left|\Psi_{in}\right\rangle$ is first encoded in the polarisation of a photon. Quantum information is then coherently transferred from $\left|\Psi_{in}\right\rangle$ into the spin-state $\left|\Psi_{X}(t)\right\rangle$ of the exciton which rotates around the equatorial plan of the Bloch sphere at an angular velocity $\left|s\right|/\hbar$. Finally, the electron-hole pair recombines and emits a photon of polarisation $\left|\Psi_{out}\right\rangle$ depending on the time spent in the solid-state. (b) Complete characterisation of the time evolution of $\left|\Psi_{X}(t)\right\rangle$ prepared in the $\left|X_{D}\right\rangle$ spin-state by time-resolved measurements in linear, diagonal, and circular bases. Each signal has been normalised by the sum of the two measurements in the corresponding basis (vertical scale from 0 to 1).  The symbols refer to measurements along different polarisations. (c) Fidelity $f_{in}$=$|\left\langle\Psi_{out}|\Psi_{in}\right\rangle|^{2}$ of the interface. The symbols refer to different input qubits.\label{fig2}}
\end{figure}

Electrical coherent control of the phase accumulated by the qubit when initialised in a superposition of eigenstates can be demonstrated by dynamic modulation of $\left|s\right|$. We operate at values $\left|s\right|>$5$\mu$eV so that the orientation of the eigenstates does not change significantly with electric field. The time evolution is then given by
\begin{equation}
\left|\Psi_{X}(t)\right\rangle=\cos\theta\left|X_{H}\right\rangle+ e^{i\frac{\int^{t}_{0}s(\tau)d\tau}{\hbar}}
e^{i\varphi}\sin\theta\left|X_{V}\right\rangle,
\end{equation}
where $s(\tau)$ is modulated by an electrical pulse inducing faster phase accumulation during the gate operation (Fig. 3(a)). By applying a pulse of constant width but different amplitudes, we show that the phase-shift is proportional to the gate amplitude (Fig. 3(b)). Some deviation from linear behaviour arises from ringing in the electrical signal due to limited speed of our device and limited bandwidth of the pulse generator. Phase-shifts were fitted from time-resolved measurements shown in Fig. 3(c) where the exciton was initialised in a diagonal superposition $\left|X_{D}\right\rangle$ and emission was measured both along the diagonal and anti-diagonal polarisations. A 500ps gaussian electrical pulse was applied 250ps after the laser initialises the qubit. The intensity falls during the gate operation as emission is Stark-shifted out of the detection window. Fig. 3(d) and 3(e) show measurements in the diagonal basis after initialisation in $\left|X_{D}\right\rangle$, respectively without gate and with a $\pi$ phase-shift gate applied, the latter corresponding to a modulation of $\left|s\right|$ from a minimum of $\approx$5$\mu$eV to a maximum of $\approx$10$\mu$eV, estimated from the Stark shift resulting from the electrical pulse. The fidelity $f_{G}$ of a gate operation is given by \cite{zoller} $f_{G}=\left|\left\langle\Psi_{G}|U_{I}|\Psi_{in}\right\rangle\right|^{2}/f_{in}$, where $\Psi_{G}$ is the state of the photon emitted by the exciton after the gate operation and $U_{I}$ is the ideal gate transformation matrix. Normalisation to $f_{in}$ is used to evaluate the fidelity of the gate independently of the initialisation of the spin-state discussed previously. To take dephasing into account, the definition can be extended to \cite{li} $f_{G}=Tr\left[\rho_{G}U_{I}\rho_{X}U^{\ast}_{I}\right]/Tr\left[\rho^{2}_{X}\right]$, where $\rho_{G}=\left|\Psi_{G}\right\rangle\left\langle\Psi_{G}\right|$ is the density matrix of the photon emitted by the exciton after the gate operation, $\rho_{X}=\left|\Psi_{X}\right\rangle\left\langle\Psi_{X}\right|$ is the density matrix of the exciton state with no gate applied, and $Tr$ represents the trace. Fidelity as a function of time for the phase-shift gate is shown in Fig. 3(f). Irregularities in the evolution of the fidelity just after the gate are due to ringing in the electrical signal. Away from the gate, the fidelity fluctuates between 0.9 and 1.

\begin{figure}
\includegraphics[width=1.\columnwidth,keepaspectratio]{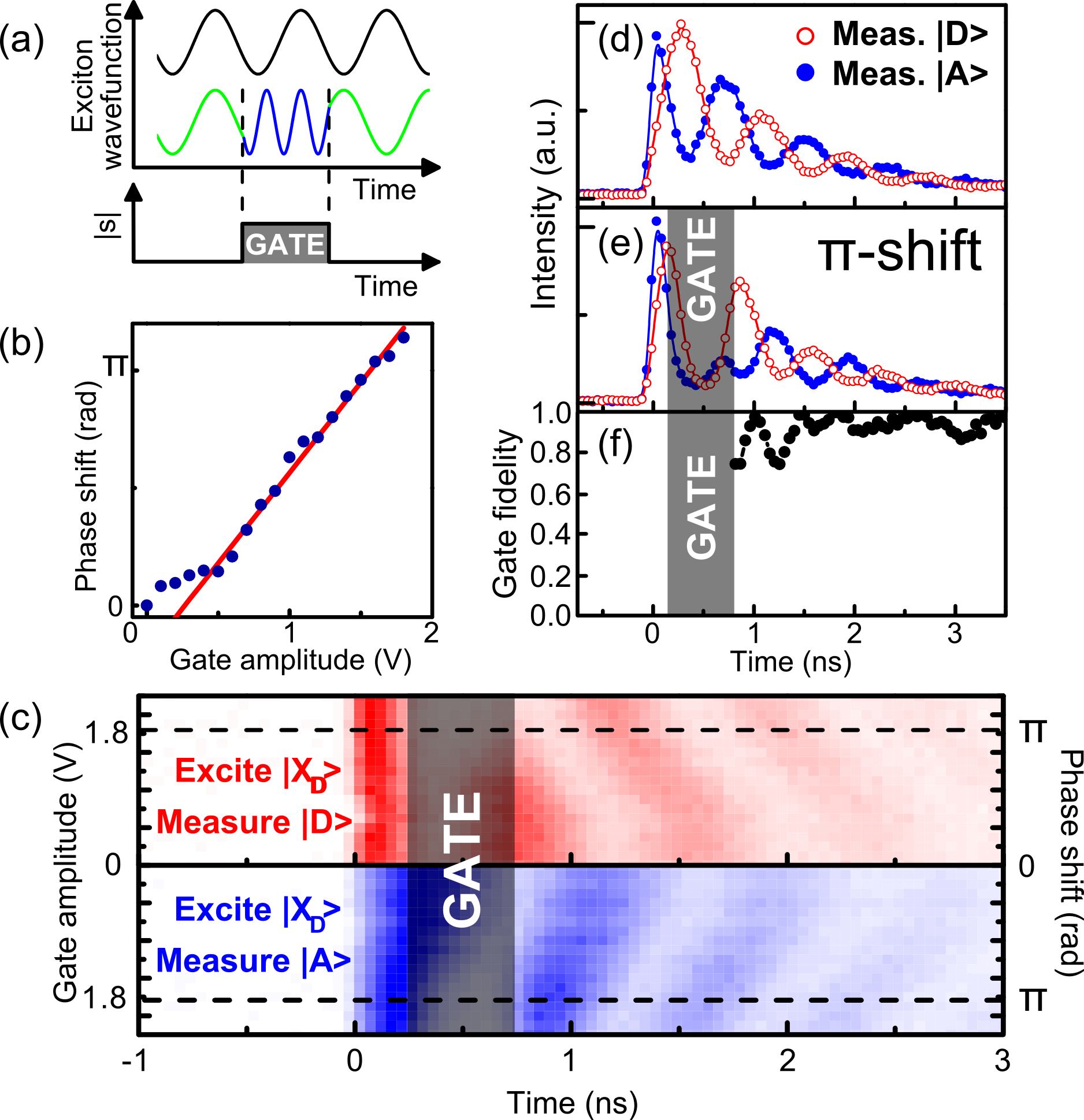}%
\caption{(colour online). (a) The electrical gate modulates $\left|s\right|$ and induces a faster phase accumulation between the spin eigenstates. The gated time-evolution (bottom) is consequently dephased compared to the ungated operation (top). (b) Phase-shift as a function of gate amplitude. (c) Normalised (colour scale from 0 to 1) time-resolved data for an exciton initialised in a diagonal superposition $\left|X_{D}\right\rangle$ and measurements in the diagonal basis. (d) and (e) Radiative emission from the QD initialised in $\left|X_{D}\right\rangle$ respectively without gate and with a $\pi$ phase-shift gate. (f) Fidelity of the $\pi$-shift gate operation as a function of time.\label{fig3}}
\end{figure}

More complex operations are achieved by operating at small values of $\left|s\right|$ where the orientation of the eigenstates are strongly sensitive to the electric field (Fig. 1(c)). This ability to change not only the energetic splitting of the two levels but also to dynamically vary the eigenstates allows complete control of the qubit: any input state can be mapped onto any output state. To demonstrate this we now describe a "spin flip" between the two eigenstates, the operating principle of which is shown in Fig. 4(a). The exciton is initialised in the eigenstate $\left|X_{D}\right\rangle$ at minimum splitting $\left|s\right|$=0.4$\mu$eV and is represented by a point at the north pole of the Bloch sphere with the axis labelled $D$. An electrical pulse similar to that used to obtain the $\pi$ phase-shift is applied to the structure and rotates the eigenstates by 45 degrees without changing the orientation of the spin-state. As a consequence, the exciton is now in a superposition $\left|X_{D}\right\rangle$ represented at the equator of the Bloch sphere and still oriented diagonally in the lab-frame. A $\pi$ phase is then accumulated during the gate operation leading to the superposition $\left|X_{A}\right\rangle$. After the pulse, the eigenstates rotate back to their initial pre-pulse orientation with the exciton being now in its eigenstate $\left|X_{A}\right\rangle$, corresponding to a spin flip from $\left|X_{D}\right\rangle$. Fig. 4(b) and 4(c) show emission from both eigenstates (oriented along $D$ and $A$), after exciting $\left|X_{D}\right\rangle$. The polarisation of the emission after the gate is reversed, corresponding to a spin-flip in the solid-state. Fig. 4(d) shows the fidelity as a function of time obtained for the spin-flip operation shown in Fig. 4(c). The fidelity is 0.97 after the gate and then fluctuates due to ringing in the electrical signal.

\begin{figure}
\includegraphics[width=1.\columnwidth,keepaspectratio]{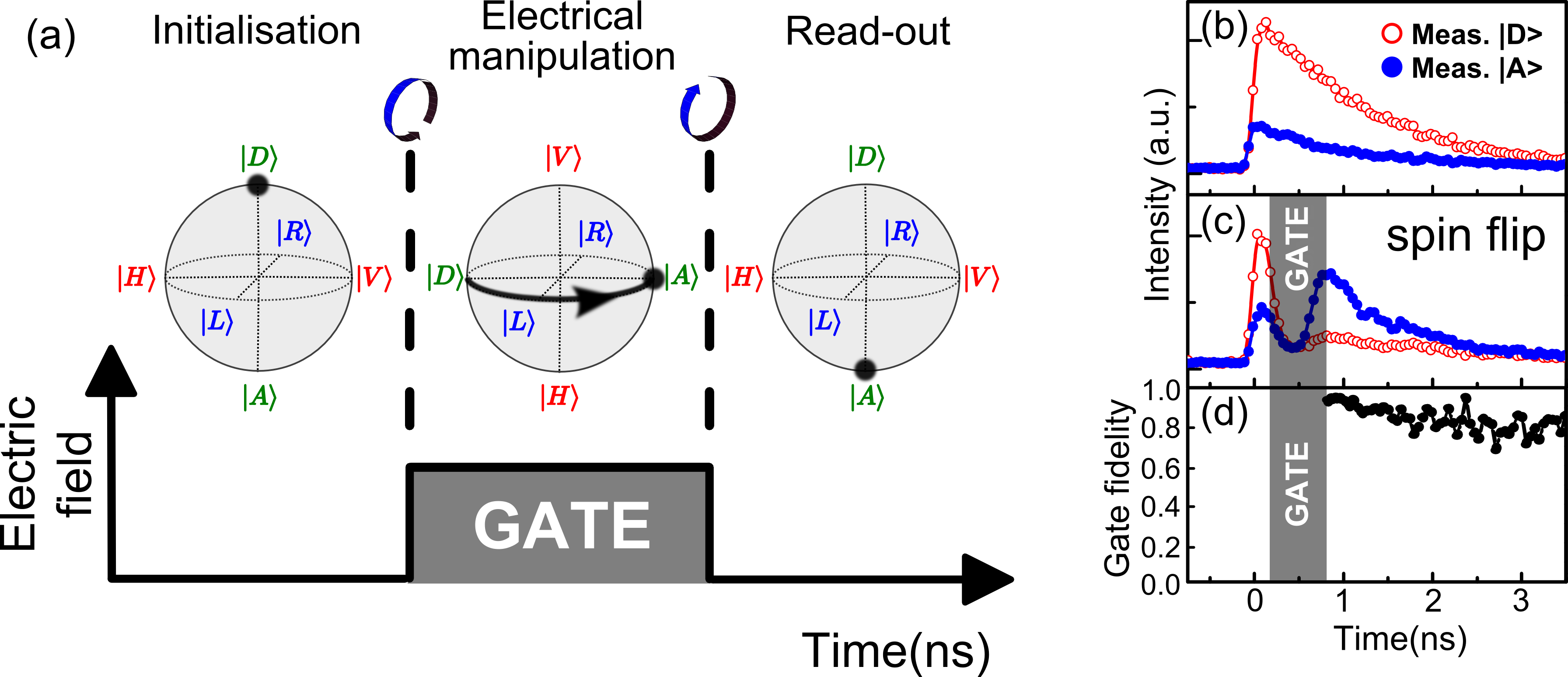}%
\caption{(colour online). (a) Operating principle of the spin-flip gate. The exciton is initialised at minimum splitting in its eigenstate $\left|X_{D}\right\rangle$. Modulating the electric field changes the value of $\left|s\right|$ and rotates the eigenstates by 45 degrees resulting in a superposition $\left|X_{D}\right\rangle$. A $\pi$ phase-shift is accumulated during the gate operation leading to a superposition $\left|X_{A}\right\rangle$. Returning to the initial value of the electric field induces another 45 degrees rotation of the eigenstates resulting in a spin flip. (b) and (c) Radiative emission from the QD respectively without gate and with the spin-flip gate. The trace with empty (plain) circles is obtained when exciting and measuring along the same (opposite) eigenstates. (d) Fidelity of the spin-flip gate as a function of time.\label{fig4}}
\end{figure}

In conclusion, we demonstrated that an exciton in a QD is a high-fidelity solid-state photonic interface, on which arbitrary phase-shifts and spin-flips can be performed electrically. Quantum information encoded in the polarisation of a photon was transfered into and restored from the spin-state of the exciton. Coherent manipulation of the spin qubits was achieved through dynamical modulation of a vertical electric field and high-fidelity phase-shift and spin-flip gate operations were demonstrated. Using small gate areas and on-chip electronics will reduce ringing in the electrical signals and improve the gate fidelities. It will also allow operation times below 10ps \cite{chau}, so that more than 300 operations can be performed within the coherence time. Furthermore, reducing the local optical density of states with cavity QED \cite{purcell} or using quantum dot molecules to separate the electron-hole pair and prevent recombination \cite{michler} would increase the exciton's lifetime by orders of magnitude and allow for triggered emission. Extending our scheme to two qubits could be achieved in the near term through the creation of biexcitons \cite{li}. Further increase in the number of qubits would be achievable using site-positioned dots with low fine-structure splitting in devices with local gates, which is technologically feasible \cite{mohan}.

\begin{acknowledgments}
We thank the EU FP7 FET programme for partial support through the Q-ESSENCE Integrated Project. Some of the authors (A.B.d.l.G, M.A.P. and R.B.P.) would also like to thank EPSRC and TREL for financial support.
\end{acknowledgments}

\end{document}